\documentclass{article}
\usepackage{spconf}
\usepackage{amsmath,amsthm,graphicx}
\usepackage{float}
\usepackage{textcomp}
\usepackage{gensymb}
\usepackage{algorithm}
\usepackage{algpseudocode}
\usepackage{amssymb,xcolor}
\usepackage{cite}
\usepackage[normalem]{ulem}
\graphicspath{{./Figs/}}

\newcommand{\argmax}{\mathop{\rm argmax}}

\renewcommand{\P}{\mathbb{P}}

\makeatother

\title{Network Infection Source Identification Under the SIRI Model}
\name{Wuhua Hu, Wee Peng Tay, Athul Harilal, and Gaoxi Xiao}
\address{School of Electrical and Electronic Engineering, Nanyang Technological University, Singapore}

\begin{document}
\ninept
\maketitle
\begin{abstract}
We study the problem of identifying a single infection source in a network under the susceptible-infected-recovered-infected (SIRI) model. We describe the infection model via a state-space model, and utilizing a state propagation approach, we derive an algorithm known as the heterogeneous infection spreading source (HISS) estimator, to infer the infection source. The HISS estimator uses the observations of node states at a particular time, where the elapsed time from the start of the infection is unknown. It is able to incorporate side information (if any) of the observed states of a subset of nodes at different times, and of the prior probability of each infected or recovered node to be the infection source. Simulation results suggest that the HISS estimator outperforms the dynamic message passing and Jordan center estimators over a wide range of infection and reinfection rates.
\end{abstract}
\section{Introduction}

Consider an infection, which can be a computer virus, disease or rumor, spreading in a network of nodes. A node is said to be in infected state if it ``possesses'' that infection \cite{shah2011rumors,luo2013identifying}. For example, in the case of a rumor spreading in an online social network like Facebook, an infected node is a user who has posted the rumor on his social page in the recent past. A node is in susceptible state if it has never been infected before, or in recovered state if it has recovered from an infection. In the Facebook example, a recovered node corresponds to a user who has removed the rumor post or the post is not within a predefined number of most recent postings of the user. An infection follows a susceptible-infected (SI) model if a susceptible node may become infected if it has infected neighbors, and an infected node stays infected \cite{keeling2005networks}. The infection has a susceptible-infected-recovered (SIR) model if an infected node may recover from an infection but then stays uninfected forever \cite{keeling2005networks}, and a susceptible-infected-recovered-infected (SIRI) model if a recovered node may again relapse into an infected state, i.e., it does not require any infected neighbors to reinfect it \cite{tudor1990deterministic,van2007modeling}. In the Facebook example, a user may repost a rumor after he has removed it due to influences external to the Facebook network \cite{luo2014identify, Luo2013ICASSP}. The SIRI model reduces to an SI or SIR model if the probability of recovery or reinfection is equal to zero, respectively.

Suppose that after an unknown elapsed amount of time since the start of an infection spreading, we have a snapshot of the states of a subset of the network, and we want to identify the infection source based on this snapshot and the network topology. This is known as the network source identification problem, and has been extensively studied under the SI and SIR models. Various source estimators such as the distance (or rumor) center \cite{shah2011rumors,luo2013identifying,dong2013rooting}, Jordan center \cite{zhu2013information,zhu2014robust,luo2014identify},
dynamic message passing (DMP) estimator \cite{lokhov2013inferring,altarelli2014bayesian},
belief propagation (BP) estimator \cite{altarelli2014bayesian,altarelli2014zero}, have been proposed and studied. Each of these estimators seeks to find an approximate maximum likelihood, maximum a posteriori (MAP), or most likely infection path estimator of the true infection source, and may require different levels of a priori information about the spreading process. For example, the distance and Jordan center estimators do not require any knowledge of the infection rates, which are utilized by DMP and BP estimators.

In this paper, we consider identifying a single infection source in a network under the SIRI model, which is more general than the SI and SIR models. The SIRI model is frequently used to describe the transmission of a contagious disease with relapse, such as bovine tuberculosis or human herpes virus, in which recovered individuals may revert back to the infectious class due to reactivation of the latent infection or incomplete treatment \cite{tudor1990deterministic,van2007modeling,guo2014dynamical}. The model is also used as a simplified version of general multi-strain models, where after an initial infection, immunity against one strain only gives partial immunity against a genetically close mutant strain \cite{martins2009scaling}. A further example of SIRI type of infection spreading is rumor spreading in an online social network, as alluded to earlier in the Facebook example. It is thus of both practical and theoretical interest to consider infection spreading and source identification in a network under the SIRI model. However, the problem is more challenging than the one under the SI and SIR models because a node may become infected and recovered multiple times. Therefore, it is unclear if the infection source estimators currently proposed in the literature can be applied directly to the SIRI model, and if this will lead to significant performance deterioration.

In this paper, we aim to find an approximate MAP estimator for inferring the infection source under the SIRI model. Our estimator is derived as a non-trivial extension of the DMP estimator, which cannot be applied directly to the SIRI model because it violates the assumption of unidirectional state transitions \cite{lokhov2014dynamic}. Our estimator is also related to the revised version of DMP called DMPr in \cite{altarelli2014zero}, which is also applicable only to the SIR model. Furthermore, our new estimator is able to incorporate side information such as the prior probability of each candidate node being the infection source, and additional observations on subsets of nodes in periods other than the snapshot time. We call our new estimator the heterogeneous infection spreading source (HISS) estimator for its applicability to a network with nodes following different (i.e., SI, SIR and SIRI) infection models. Simulations are performed on random regular tree networks and a subset of Facebook network to evaluate the proposed estimator and compare its performance with those of the Jordan center estimator, and the DMP estimator. Our simulation results suggest that the HISS estimator outperforms both the Jordan center and DMP estimator over a wide range of infection and reinfection rates.

\section{Infection model and assumptions}

In this section, we characterize the SIRI infection spreading using a state-space approach \cite{Chen1999}. Throughout this paper, we assume a common underlying probability space with probability measure $\P$. We also use $0:n$ to denote the integer set $\{0,1,...,n\}$.

Let the network over which the infection spreads be described by a directed graph, $(\mathcal{N},\mathcal{E})$, where $\mathcal{N}\triangleq1:N$ is the node set and $\mathcal{E}$ is the edge set. A directed edge $(l,k)$ exists from node $l$ to node $k$ if node $l$ can directly infect node $k$, in which case node $l$ is said to be an \textit{in-neighbor} of node $k$, and conversely node $k$ is said to be an \textit{out-neighbor} of node $l$. We denote the set of in-neighbors of a node $k$ as $\mathcal{N}_{k}$.

\begin{figure}[!t]
\noindent \begin{centering}
\includegraphics[scale=0.55]{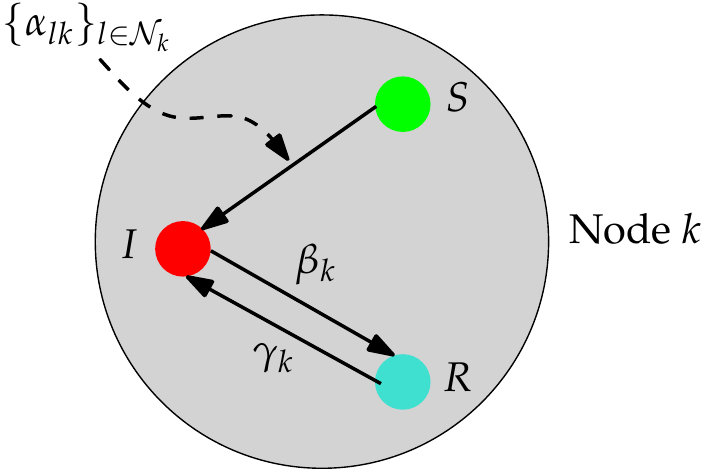}
\par\end{centering}
\protect\caption{Possible state transitions of a node under the SIRI model. \label{fig: Possible-state-transitions}}
\end{figure}

Suppose that there is a single node $s^*$ in the network that starts the infection at time 0, and suppose that we observe the node states of a set of nodes in the network at a particular time $T_f$, which we call the \emph{snapshot} time. We assume that $T_f$ is unknown, and that time is discretized into $0:T_{f}$. Let $S$, $I$, and $R$ denote the susceptible, infected, and recovered states, respectively. Let $\alpha_{lk}$ be the probability of an infected node $l$ infecting an out-neighbor $k$ within a time slot, where $\alpha_{lk}=0$ if $l\notin\mathcal{N}_{k}$. Let $\beta_{k}$ be the probability of an infected node $k$ to recover within a time slot, and $\gamma_{k}$ be the probability of a recovered node $k$ to be infected again within a time slot, where $\gamma_{k}=0$ if $\beta_{k}=0$. These probabilities are assumed to be given or have been inferred a priori. The possible state transitions of a node are depicted in Fig. \ref{fig: Possible-state-transitions}. It is evident that the infection process at node $k$ reduces to the SIR model if $\gamma_{k}=0$, and the SI model if we further have $\beta_{k}=0$. This admits a heterogeneous spreading model containing different infection processes of SI, SIR and SIRI at different nodes. Our model thus subsumes those studied in \cite{shah2011rumors,luo2013identifying,zhu2013information,lokhov2014dynamic,altarelli2014zero}. We assume that conditioned on the node states in time slot $t-1$, all node transitions in time slot $t$ are independent of each other.

We use $P_{k}^{S}(t)$, $P_{k}^{I}(t)$ and $P_{k}^{R}(t)$ to denote
the probabilities of a node $k\in\mathcal{N}$ to be in states $S$,
$I$ and $R$ at time $t$, respectively. Adopting a state-space modeling
approach, we define these three probabilities as state variables.
The infection model is then obtained by describing the evolution of
the three state variables in time using difference equations. Using the state transitions for the SIRI model, we have for all $t\in1:T_{f}$ and $k\in\mathcal{N}$,
\begin{align}
P_{k}^{S}(t) &= \P\left(\bar{U}_{k}(t-1)\mid S_{k}(t-1)\right)\cdot P_{k}^{S}(t-1),\label{eq: S-dynamics}\\
P_{k}^{R}(t) &= \beta_{k}P_{k}^{I}(t-1)+(1-\gamma_{k})P_{k}^{R}(t-1),\label{eq: I-dynamics}\\
P_{k}^{I}(t) &= 1-P_{k}^{S}(t)-P_{k}^{R}(t),\label{eq: R-dynamics}
\end{align}
where $\bar{U}_{k}(t-1)$ is the event that no in-neighbor passes the infection
to node $k$ in during time $t-1$ to $t$, and $S_{k}(t-1)$
is the event that node $k$ is in state $S$ at time $t-1$. By assuming
that in-neighbors pass infection to a node independently, and utilizing the mean-field approximation, we have 
\begin{equation}
\mathbb{P}\left(\bar{U}_{k}(t-1)\mid S_{k}(t-1)\right)\approx \prod_{l\in\mathcal{N}_{k}} \left(1-\alpha_{lk}P_{l}^{I\mid S_{k}(t-1)}(t-1) \right),\label{eq: key conditional probability}
\end{equation}
where $P_{l}^{I\mid S_{k}(t_{1})}(t_{2})\triangleq\mathbb{P}\left(I_{l}(t_{2})\mid S_{k}(t_{1})\right)$
, for all $t_{1},t_{2}\in 0:T_{f}$, which is the probability that
node $l$ is in state ${\it I}$ at time $t_{2}$ given that node
$k$ is in state ${\it S}$ at time $t_{1}$. (Note that equation \eqref{eq: key conditional probability} is exact if the graph $(\mathcal{N},\mathcal{E})$ is acyclic.) To complete the model,
we need to have an explicit expression for $P_{l}^{I\mid S_{k}(t-1)}(t-1)$.

To that end, we introduce two classes of auxiliary state variables.
For all $t\in0:T_{f},l\in\mathcal{N}_{k},k\in\mathcal{N}$, let $\theta_{lk}(t)$
be the the probability of a node $l$ not infecting its out-neighbor $k$ up to time $t$, given that node $k$ is in state $S$ at time $t$; and let $\phi_{lk}(t)$ be the probability of a node
$l$ to be in state $I$ at time $t$ and not infecting its out-neighbor $k$ up to time $t$, given that node $k$ is in state $S$ at time $t$. Then it can be shown that
\begin{equation}
P_{l}^{I\mid S_{k}(t-1)}(t-1)=\dfrac{\phi_{lk}(t-1)}{\theta_{lk}(t-1)},\label{eq: conditioned-I-dynamics}
\end{equation}
where $\theta_{lk}(t-1)$ and $\phi_{lk}(t-1)$ are updated by
\begin{align}
\theta_{lk}(t)= & \theta_{lk}(t-1)-\alpha_{lk}\phi_{lk}(t-1),\label{eq: theta-update}\\
\phi_{lk}(t)= & (1-\alpha_{lk})(1-\beta_{l})\phi_{lk}(t-1)\nonumber \\
 & +\left(P_{l}^{S\mid S_{k}(t-1)}(t-1)-P_{l}^{S\mid S_{k}(t)}(t)\right)\nonumber \\
 & +\gamma_{l}P_{l}^{R\mid S_{k}(t-1)}(t-1),\label{eq: phi-update}
\end{align}
and the auxiliary conditional probabilities are computed from
\begin{align}
P_{l}^{S\mid S_{k}(t-1)}(t-1)= & P_{l}^{S}(0)\prod_{j\in\mathcal{N}_{l}\backslash\{k\}}\theta_{jl}(t-1),\label{eq: conditional P_S undate}\\
P_{l}^{R\mid S_{k}(t-1)}(t-1)= & 1-P_{l}^{S\mid S_{k}(t-1)}(t-1)-P_{l}^{I\mid S_{k}(t-1)}(t-1).\label{eq: conditional P_R update}
\end{align}
Compared to the DMP equations in \cite{lokhov2013inferring}, the new term at the end of \eqref{eq: phi-update} arises from the reinfection of a recovered node, which is unique
to the SIRI infection model. This term is then computed by (\ref{eq: conditional P_R update}),
which further depends on the new equation (\ref{eq: conditioned-I-dynamics})
and equation (\ref{eq: conditional P_S undate}). The
key equation (\ref{eq: conditioned-I-dynamics}) follows from an important
observation:
\begin{align*}
P_{k}^{S}(t) & =P_{k}^{S}(0)\prod_{l\in\mathcal{N}_{k}}\theta_{lk}(t)=P_{k}^{S}(t-1)\prod_{l\in\mathcal{N}_{k}}\dfrac{\theta_{lk}(t)}{\theta_{lk}(t-1)}\\
 & =P_{k}^{S}(t-1)\prod_{l\in\mathcal{N}_{k}}\left(1-\alpha_{lk}\dfrac{\phi_{lk}(t-1)}{\theta_{lk}(t-1)}\right)\\
 & = P_{k}^{S}(t-1)\prod_{l\in\mathcal{N}_{k}}\left(1-\alpha_{lk}P_{l}^{I\mid S_{k}(t-1)}(t-1)\right),
\end{align*}
where the last equality follows from \eqref{eq: S-dynamics} and \eqref{eq: key conditional probability}. We note that for a directed acyclic graph, the above equations are exact, while we will treat these as approximations for general network graphs.

We proceed to associate \textit{virtual} observation equations with
the state dynamics. Let $\mathcal{N}_{t}^{S}$, $\mathcal{N}_{t}^{I}$ and
$\mathcal{N}_{t}^{R}$ denote the sets of nodes observed to be in
states $S$, $I$ and $R$ at time $t$, respectively; and $\mathcal{N}_{t}^{SR}$
denote the set of nodes observed to be in an uninfected state but are
indistinguishable to be in state $S$ or $R$ at time $t$. The rest of the
nodes whose states are unknown or not observed are collected into
a set denoted by $\mathcal{N}_{t}^{SIR}$, i.e., $\mathcal{N}_{t}^{SIR}=\mathcal{N}\backslash(\mathcal{N}_{t}^{S}\cup\mathcal{N}_{t}^{I}\cup\mathcal{N}_{t}^{R}\cup\mathcal{N}_{t}^{SR})$.
Stack the three states into a single vector as
$P_{k}(t)\triangleq[P_{k}^{S}(t)\, P_{k}^{I}(t)\, P_{k}^{R}(t)]^{T}$.
We define the virtual observations as
\begin{equation}
y_{k}(t)=C_{k}(t)P_{k}(t),\label{eq: observation}
\end{equation}
where the observation vector $C_{k}(t)\in\mathbb{R}^{1\times3}$ takes one of the feasible values:
$[1\,\, 0\,\, 0]$, if $k\in\mathcal{N}_{t}^{S}$;
$[0\,\, 1\,\, 0]$, if $k\in\mathcal{N}_{t}^{I}$;
$[0\,\, 0\,\, 1]$, if $k\in\mathcal{N}_{t}^{R}$;
$[1\,\, 0\,\, 1]$, if $k\in\mathcal{N}_{t}^{SR}$;
$[1\,\, 1\,\, 1]$, if $k\in\mathcal{N}_{t}^{SIR}$.
The last observation vector corresponds to a null observation and
is defined merely for theoretical completeness.


\section{Identifying the infection source}

We assume that at a particular observation snapshot time $T_f$, we wish to infer the identity of the infection source. For each node $k$, let $W_{k}$ be the set of time slots during which the state of node $k$ is observed. The time slots in $W_k$ are known only relative to the unknown snapshot time $T_f$, which is to be inferred. For each $t \in W_{k}$, let $X_{k,t} \in\{S,I,R,SR\}$ be the observed state of node $k$ at time $t$, where $SR$ represents an observation state in which we are unable to distinguish an uninfected state as either $S$ or $R$. We lump all observation times into a set $\mathcal{T}^{O} =\cup_{k\in\mathcal{N}}W_{k}$. Let $\mathcal{N}^{O}$ be the set of all nodes with at least one observation up to the snapshot time. 

We exclude all observed susceptible nodes whose in-neighbors are also observed to be susceptible, because these nodes do not contribute to the infection realization \cite{altarelli2014bayesian}. We use $\mathcal{N}'$ to denote the full node set $\mathcal{N}$ with such
uninformative nodes removed, and $\mathcal{N}_{k}'$ to denote the neighbors of node $k$ in the reduced graph for all $k\in\mathcal{N}'$. 

Let $\mathcal{N}^{c}$ be the set of nodes that have been observed to be in states $I$, $R$, or $SR$ at some observation time (i.e., the candidate sources).
Moreover, the prior probability of a candidate to be the infection source
is denoted by $P_{k,0}^{I}$ for all $k\in\mathcal{N}^{c}$.

We formulate the source identification problem as an approximate MAP estimation of the infection source. Ideally, the true infection source and the true snapshot time should be estimated as
\begin{equation}
(\hat{s},\hat{T_{f}})\in\argmax_{s\in\mathcal{N}^{c},T_{f}\in\mathcal{T}^{c}}\P\left(s^*=s,\mathcal{N}^{O}\right),\label{eq: ideal estimator}
\end{equation}
where $\mathcal{T}^{c}$ is the set of candidate snapshot times, and
$\P\left(s^*=s,\mathcal{N}^{O}\right)$ is the joint probability of node $s$ being the infection source $s^*$, and $\mathcal{N}^{O}$ being the given observations. By Bayes' rule, the ideal estimator is equivalent to maximizing  $\mathbb{P}\left(\mathcal{N}^{O}\mid s^*=s\right)\mathbb{P}(s^*=s)$. For tractability, we approximate the joint probability $\mathbb{P}\left(\mathcal{N}^{O}\mid s^*=s\right)$
by a mean-field probability $\prod_{k\in\mathcal{N}^{O}}\prod_{t \in W_{k}}P_{k}^{X_{k, t}\mid I_{s}(0)}(t)$. As the probabilities $P_{k}^{X_{k,i}\mid I_{s}(0)}(t)$ can be computed by iterating the infection model (\ref{eq: S-dynamics})-(\ref{eq: observation}) for a given infection source $s$, this mean-field probability can be computed and then used to compare different candidate infection
sources. Together with an approximate prior $\mathbb{P}(s^*=s)$, the estimator gives an approximate MAP estimation of the infection source.

Therefore, the inference problem can be written in the following optimization form:
\begin{align}
(\text{P0})\max_{T_{f},\{P_{k}^{I}(0)\}_{k\in\mathcal{N}^{c}}} & \sum_{s\in\mathcal{N}^{c}}
P_{s}^{I}(0) \prod_{k\in\mathcal{N}^{O}}\prod_{t \in W_{k}} y_{k}(t)\nonumber \\
\text{subject to} & \quad \eqref{eq: S-dynamics}-\eqref{eq: observation},\nonumber \\
 & P_{k}^{S}(0)=1-P_{k}^{I}(0),\forall k\in\mathcal{N}^{c},\label{eq: initial-conds-start}\\
 & P_{k}^{S}(0)=1,\forall k\in\mathcal{N}'\backslash\mathcal{N}^{c},\\
 & P_{k}^{I}(0)=0,\forall k\in\mathcal{N}'\backslash\mathcal{N}^{c},\\
 & P_{k}^{R}(0)=0,\forall k\in\mathcal{N}',\\
 & \theta_{lk}(0)=1,\forall l\in\mathcal{N}_{k}',k\in\mathcal{N}',\\
 & \phi_{lk}(0)=P_{k}^{I}(0),\forall l\in\mathcal{N}_{k}',k\in\mathcal{N}',\label{eq: initial-conds-end}\\
 & \sum_{k\in\mathcal{N}^{c}}P_{k}^{I}(0)=1,\nonumber \\
 & T_{f}\in\mathcal{T}^{c},P_{k}^{I}(0)\in\{0,1\},\forall k\in\mathcal{N}^{c},\nonumber
\end{align}
where the infection model \eqref{eq: S-dynamics}-\eqref{eq: observation}
is applied to a reduced graph which excludes the aforementioned observed uninformative
susceptible nodes, and (\ref{eq: initial-conds-start})-(\ref{eq: initial-conds-end})
specify the initial conditions of the infection model. 

Note that the observation
variables $y_{k}(t)$ in the objective function are expressed by state
variables $P_{k}(t)$ via (\ref{eq: observation}), and that the basic
state variables $P_{k}(t)$ and the auxiliary state variables $\theta_{lk}(t)$
and $\phi_{lk}(t)$, for all $t\ge0$, $l\in\mathcal{N}_{k}$ and
$k\in\mathcal{N}$, are determined by the decision variables $P_{k}^{I}(0)$
for all $k\in\mathcal{N}^{c}$ via (\ref{eq: S-dynamics})-(\ref{eq: observation}).
Therefore, the optimization essentially depends on the decision variables $\{P_{k}^{I}(0)\}_{k\in\mathcal{N}^{c}}$ and $T_{f}$ . We also note that $T_{f}$ is larger than the elapsed time between the time when the first side information is observed and the snapshot time.

Because of the nonlinearities involved in the
infection model (\ref{eq: S-dynamics})-(\ref{eq: observation}),
it is difficult to solve (P0) by a standard optimization solver. Nonetheless,
it is straightforward to get a solution by enumerating and comparing
all feasible solutions of (P0). Given a candidate source and a value
of $T_{f}$, the algorithmic complexity of computing the objective
function is $O(dT_{f}|\mathcal{N}'|)$, where $d$ is the average
in-degree of the reduced graph. Since we need to enumerate all candidate
values of $T_{f}$ for all candidate sources, the overall algorithmic
complexity is then $O(dT_{f}|\mathcal{N}'||\mathcal{N}^{c}||\mathcal{T}^{c}|)$.

Our proposed inference process inherits the merit of the DMP method, but it is able to incorporate available side information (if any) and is applicable to a network consisting of nodes that follow heterogeneous (namely, SI, SIR and SIRI ) infection models. For this reason, we call this new estimator the HISS estimator for short.

Although the HISS estimator is derived by assuming that the in-neighbors
pass infections to a node independently (cf. (\ref{eq: key conditional probability})), we can still apply it heuristically to a network where this assumption is violated. We expect that the impact on the performance of the HISS estimator to be small if the network is sparse or has weak correlations
between infections passed by neighboring nodes. We note that a similar argument was used to explain the success of well-known BP methods in general network \cite{murphy1999loopy}.

\section{Performance evaluation via simulations}

We apply the proposed HISS estimator to infer single infection sources in random
regular tree networks and a subset of the Facebook network under the SIRI model, and then compare the inference results with those obtained with the DMP and Jordan centre (JC) estimators. In applying the DMP estimator, we ignore the reinfection probability of a recovered node and hence naively treat the SIRI model as an SIR model. In the simulations, we assume that the snapshot time falls in the range $[\tau+1,T_{f}+\tau-1]$, where $T_{f}$ is the actual elapsed time and $\tau$ is the elapsed time since the first side information is observed till the snapshot observation time. In simulations we set $\tau$ to $T_{f}/2$ if $T_{f}$ is even and $(T_{f}-1)/2$ if $T_{f}$ is odd. We assume uniform infection, recovery and reinfection rates at all nodes in the network. 

To compare the estimators, we adopt the following two metrics, which are correlated to the likelihood of wrongly inferring the source and the absolute inference error in distance, respectively: 
\begin{itemize}
	\item \emph{Normalized rank of $s^*$} \cite{lokhov2013inferring,altarelli2014bayesian}: We first rank the candidate sources in descending order according to the objective function value in (P0). The rank ($\ge1$) of the true infection source, subtracted by the ideal rank value of 1 (yielding the rank error), and then divided by the total number of candidate sources, gives in the desired normalized rank. 
	\item \emph{Error distance of $\hat{s}$ to $s^*$} \cite{Wuqiong2014}: We compute the length of the shortest path from the source estimate $\hat{s}$ to the true source $s^*$. 
\end{itemize}

\begin{figure}
\begin{centering}
\includegraphics[width=0.5\textwidth]{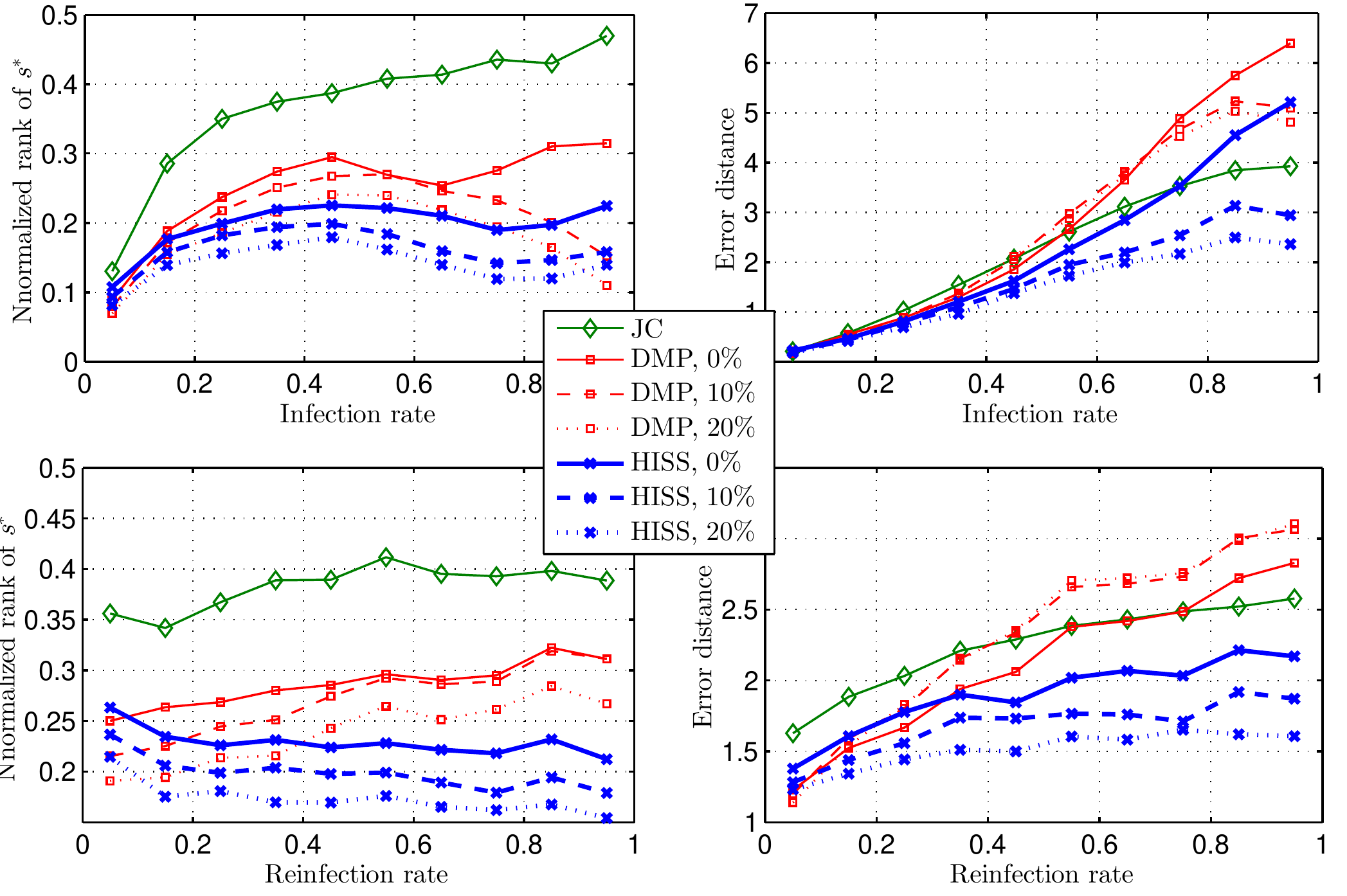}
\end{centering}
\caption{Comparison of estimators' performance for random regular tree networks. The percentage refers to the fraction of nodes, selected at random at time $T_f-\tau$, whose states are observed. \label{fig: Inference-results-on-tree}}
\end{figure}

In the case of regular tree networks, in each instance, we randomly generate
a tree network of 1000 nodes with each node's degree equal to 4 \cite{lokhov2013inferring,altarelli2014bayesian}, and randomly choose an infection source. We then simulate the infection
spreading until either 10 time slots or no less than 20\% of the nodes
are infected or recovered. We observe the states of all nodes at the last simulation time slot. To simulate the availability of different side information, we also collect the states of a random selection of a given percentage ($0\%$, $10\%$ or $20\%$) of the nodes at time $T_{f}-\tau$. The inference results, averaged over 1000 random instances, are shown in Fig. \ref{fig: Inference-results-on-tree}. As the infection rate increases while the reinfection rate is fixed at 0.5, we observe that HISS outperforms DMP and JC for most infection rates w.r.t. both performance metrics. We also observe that incorporating additional observations (side information) always improves the performance of HISS. However, this is not always true for DMP since the side information may mislead the inference process due to the spreading model mismatch. As the reinfection rate increases while the infection rate is fixed at 0.5, we observe that the advantage of HISS over DMP and JC becomes more obvious due to its use of a more general infection model in the inference. Again, incorporating additional observations improves the performance of HISS but may deteriorate that of DMP. The results also show that an estimator that is stronger in one performance metric may be weaker in the other performance metric, which is clear by comparing the performances of DMP and JC.

\begin{figure}
\begin{centering}
\includegraphics[width=0.5\textwidth]{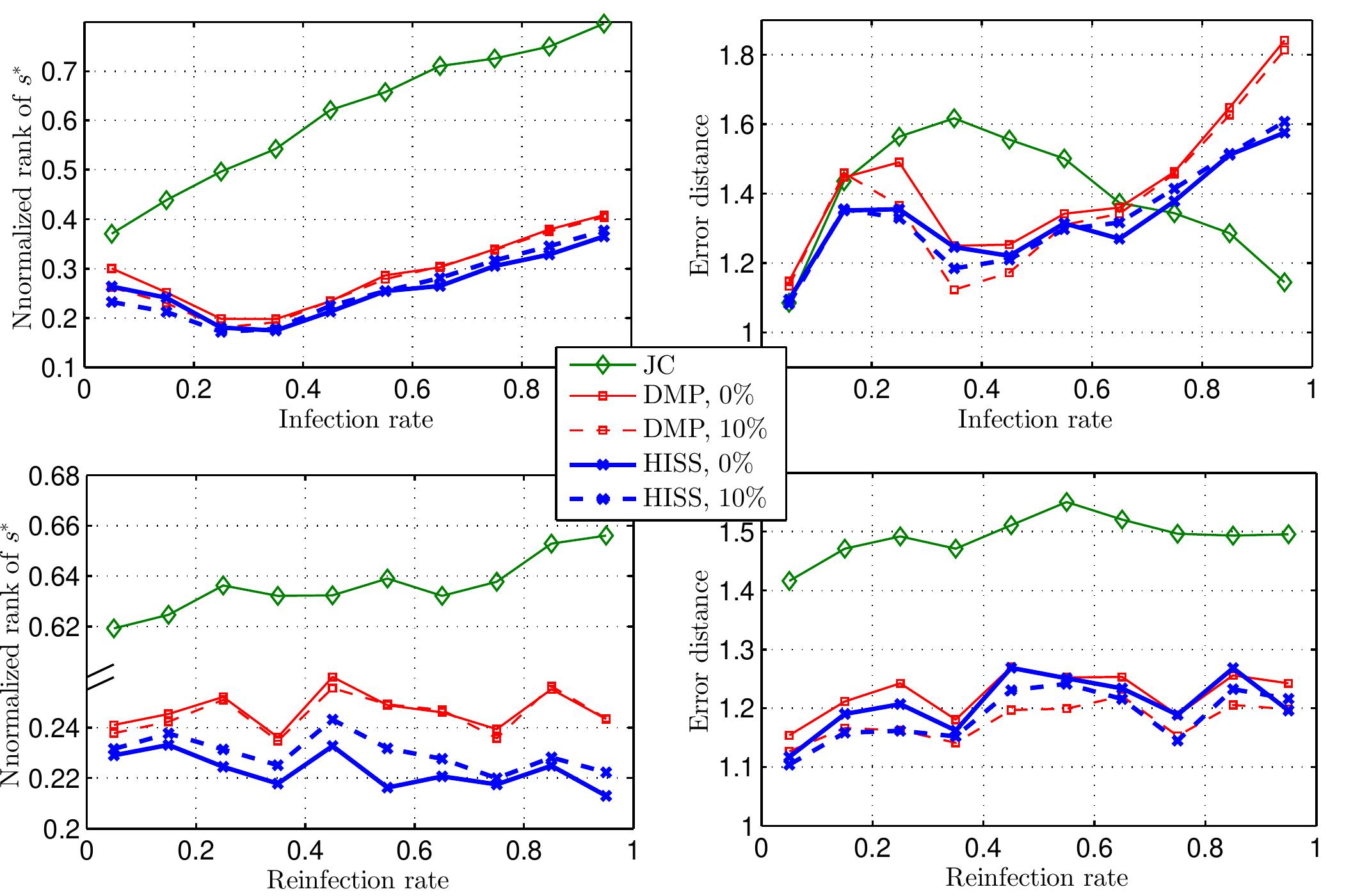}
\end{centering}
\caption{Comparison of estimators' performance for a subset of the Facebook network. The percentage refers to the fraction of nodes, selected at random at time $T_f-\tau$, whose states are observed.\label{fig: Inference-results-on-facebook}}
\end{figure}

In the case of the Facebook network, we arbitrarily select a subset of 500 nodes from the Facebook dataset used in \cite{leskovec2012learning}. In each simulation instance, we randomly specify an infection source and perform simulations under the same parameter settings as the regular trees. The results are shown in Fig. \ref{fig: Inference-results-on-facebook}. We observe that, in the absence of side information, HISS outperforms DMP w.r.t.\ both performance metrics for most infection and reinfection rates. However, incorporating additional observations does not necessarily improve and may even deteriorate the performance of both HISS and DMP, in which case HISS may become inferior to DMP. This occurs because the Facebook network subset is found to be very loopy and contain many circular subsets, which tends to invalidate the mean-field assumption embedded in the HISS and DMP estimators. Despite this intrinsic limitation, both DMP and HISS outperform JC for almost all infection and reinfection rates, except when infection rates are high in which case JC has the smallest average error distance. Its corresponding average normalized rank of the true source however remains the largest among the three estimators.

\section{Conclusion}

We have introduced a state-space description of an SIRI infection model, and using the state propagation equations, we have derived an approximate MAP estimator for the infection source, given the observations of a set of node states at a snapshot time. Our proposed estimator is able to incorporate side information like observations of node states at intermediate times during the infection spreading and prior beliefs of potential candidate infection sources, into the inference procedure. Simulations
on random regular tree networks and a subset of the Facebook network suggest that the HISS estimator outperforms the Jordan center and DMP estimator for a wide range of the infection and reinfection rates. However, we note that in networks that are very loopy, adding side information may lead to a performance deterioration. Future work includes designing better inference procedures that can handle side information better for loopy networks.

\section{Acknowledgment}

The authors are grateful to Andrey Y. Lokhov for sharing with us the
DMP implementation code which helped us in implementing the HISS method
and doing the comparisons. The authors also thank Wuqiong Luo for helpful discussions. The research is supported in part by the Singapore Ministry of Education Academic Research Fund Tier 2 grant MOE2013-T2-2-006.

\bibliographystyle{IEEEtran}
\bibliography{IEEEabrv,DMPlus}

\begin{thebibliography}{10}
\providecommand{\url}[1]{#1}
\csname url@samestyle\endcsname
\providecommand{\newblock}{\relax}
\providecommand{\bibinfo}[2]{#2}
\providecommand{\BIBentrySTDinterwordspacing}{\spaceskip=0pt\relax}
\providecommand{\BIBentryALTinterwordstretchfactor}{4}
\providecommand{\BIBentryALTinterwordspacing}{\spaceskip=\fontdimen2\font plus
\BIBentryALTinterwordstretchfactor\fontdimen3\font minus
  \fontdimen4\font\relax}
\providecommand{\BIBforeignlanguage}[2]{{%
\expandafter\ifx\csname l@#1\endcsname\relax
\typeout{** WARNING: IEEEtran.bst: No hyphenation pattern has been}%
\typeout{** loaded for the language `#1'. Using the pattern for}%
\typeout{** the default language instead.}%
\else
\language=\csname l@#1\endcsname
\fi
#2}}
\providecommand{\BIBdecl}{\relax}
\BIBdecl

\bibitem{shah2011rumors}
D.~Shah and T.~Zaman, ``Rumors in a network: Who's the culprit?'' \emph{{IEEE}
  Trans. Inf. Theory}, vol.~57, no.~8, pp. 5163--5181, 2011.

\bibitem{luo2013identifying}
W.~Luo, W.~P. Tay, and M.~Leng, ``Identifying infection sources and regions in
  large networks,'' \emph{{IEEE} Trans. Signal Process.}, vol.~61, no.~11, pp.
  2850--2865, 2013.

\bibitem{keeling2005networks}
M.~J. Keeling and K.~T. Eames, ``Networks and epidemic models,'' \emph{Journal
  of the Royal Society Interface}, vol.~2, no.~4, pp. 295--307, 2005.

\bibitem{tudor1990deterministic}
D.~Tudor, ``A deterministic model for herpes infections in human and animal
  populations,'' \emph{SIAM Review}, vol.~32, no.~1, pp. 136--139, 1990.

\bibitem{van2007modeling}
P.~Van~den Driessche and X.~Zou, ``Modeling relapse in infectious diseases,''
  \emph{Mathematical Biosciences}, vol. 207, no.~1, pp. 89--103, 2007.

\bibitem{luo2014identify}
W.~Luo, W.~Tay, and M.~Leng, ``How to identify an infection source with limited
  observations,'' \emph{{IEEE} J. Sel. Topics Signal Process.}, vol.~8, no.~4,
  pp. 586 -- 597, 2014.

\bibitem{Luo2013ICASSP}
W.~Luo and W.~P. Tay, ``Finding an infection source under the \textrm{SIS}
  model,'' in \emph{Proc. IEEE 38th International Conference on Acoustics,
  Speech, and Signal Processing (ICASSP)}, 2013.

\bibitem{dong2013rooting}
W.~Dong, W.~Zhang, and C.~W. Tan, ``Rooting out the rumor culprit from
  suspects,'' in \emph{2013 IEEE International Symposium on Information Theory
  Proceedings (ISIT)}, Istanbul, Turkey, 2013, pp. 2671--2675.

\bibitem{zhu2013information}
K.~Zhu and L.~Ying, ``Information source detection in the sir model: a sample
  path based approach,'' in \emph{Information Theory and Applications Workshop
  (ITA), 2013}, 2013, pp. 1--9.

\bibitem{zhu2014robust}
------, ``A robust information source estimator with sparse observations,''
  \emph{To appear in Computational Social Networks, arXiv preprint
  arXiv:1309.4846}, 2014.

\bibitem{lokhov2013inferring}
A.~Y. Lokhov, M.~M{\'e}zard, H.~Ohta, and L.~Zdeborov{\'a}, ``Inferring the
  origin of an epidemic with dynamic message-passing algorithm,'' \emph{Phys.
  Rev. E}, vol.~9, no. 012801, 2014.

\bibitem{altarelli2014bayesian}
F.~Altarelli, A.~Braunstein, L.~Dall'Asta, A.~Lage-Castellanos, and
  R.~Zecchina, ``Bayesian inference of epidemics on networks via belief
  propagation,'' \emph{Phys. Rev. Letts.}, vol. 112, no.~11, p. 118701, 2014.

\bibitem{altarelli2014zero}
F.~Altarelli, A.~Braunstein, L.~Dall'Asta, A.~Ingrosso, and R.~Zecchina, ``The
  zero-patient problem with noisy observations,'' \emph{arXiv preprint
  arXiv:1408.0907}, 2014.

\bibitem{guo2014dynamical}
P.~Guo, X.~Yang, and Z.~Yang, ``Dynamical behaviors of an siri epidemic model
  with nonlinear incidence and latent period,'' \emph{Advances in Difference
  Equations}, vol. 2014, no.~1, p. 164, 2014.

\bibitem{martins2009scaling}
J.~Martins, A.~Pinto, and N.~Stollenwerk, ``A scaling analysis in the siri
  epidemiological model,'' \emph{Journal of Biological Dynamics}, vol.~3,
  no.~5, pp. 479--496, 2009.

\bibitem{lokhov2014dynamic}
A.~Y. Lokhov, M.~M{\'e}zard, and L.~Zdeborov{\'a}, ``Dynamic message-passing
  equations for models with unidirectional dynamics,'' \emph{arXiv preprint
  arXiv:1407.1255}, 2014.

\bibitem{Chen1999}
C.~T. Chen, \emph{{L}inear {S}ystem {T}heory and {D}esign}.\hskip 1em plus
  0.5em minus 0.4em\relax Oxford: Oxford University Press, 1999.

\bibitem{murphy1999loopy}
K.~P. Murphy, Y.~Weiss, and M.~I. Jordan, ``Loopy belief propagation for
  approximate inference: An empirical study,'' in \emph{Proceedings of the
  Fifteenth Conference on Uncertainty in Artificial Intelligence}.\hskip 1em
  plus 0.5em minus 0.4em\relax Morgan Kaufmann Publishers Inc., 1999, pp.
  467--475.

\bibitem{Wuqiong2014}
W.~Luo, ``Identifying infection sources in a network,'' Ph.D. dissertation,
  Nanyang Technological University, Singapore, Aug 2014.

\bibitem{leskovec2012learning}
J.~Leskovec and J.~J. Mcauley, ``Learning to discover social circles in ego
  networks,'' in \emph{Advances in Neural Information Processing Systems 25},
  2012, pp. 539--547.

\end{thebibliography}

\end{document}